\documentclass[traditabstract]{aa}
\usepackage{amsmath,bm}
\usepackage{txfonts}
\usepackage{natbib}
\usepackage{graphicx}
\usepackage{multirow}
\bibpunct{(}{)}{;}{a}{}{,}

\newcommand{\be}{\begin{equation}}
\newcommand{\ee}{\end{equation}}
\newcommand{\bea}{\begin{eqnarray}}
\newcommand{\eea}{\end{eqnarray}}
\newcommand{\ie}{{\it i.e.}}
\newcommand{\eg}{{\it e.g.}}

\begin{document}

\title{Composition and stability of hybrid stars with hyperons and
  quark color-superconductivity}

\abstract{ The recent measurement of a $1.97\pm 0.04$ solar-mass
  pulsar places a stringent lower bound on the maximum mass of compact
  stars and therefore challenges the existence of any agents that
  soften the equation of state of ultra-dense matter.  We investigate
  whether hyperons and/or quark matter can be accommodated in massive
  compact stars by constructing an equation of state based on a
  combination of phenomenological relativistic hyper-nuclear density
  functional and an effective model of quantum chromodynamics (the
  Nambu-Jona-Lasinio model).  Stable configurations are obtained with
  $M \ge 1.97 M_{\sun}$ featuring hyper-nuclear and quark matter in
  color superconducting state if the equation of state of nuclear
  matter is stiff above the saturation density, the transition to
  quark matter takes place at a few times the nuclear saturation
   density, and the repulsive vector interactions in quark matter are
  substantial.  }

\author{Luca~Bonanno and Armen~Sedrakian}
\institute{Institute for Theoretical Physics, J.-W. Goethe University, 
D-60438  Frankfurt-Main, Germany}
\keywords{Neutron stars, Dense Matter, Pulsars}
\titlerunning{Hybrid stars with hyperons and quarks}
\maketitle

\section{Introduction}
The masses of neutron stars  are the most
sensitive among all their parameters to the equation of state at high densities. Therefore,
pulsar mass measurements provide one of the key experimental
constraints on the theory of ultra-dense matter.  The masses measured
in the pulsar binaries are clustered around the value 1.4~$M_{\sun}$
and have been consider as ``canonical'' for a long time.  However, in
recent years mounting evidence emerged in favor of substantially heavier
neutron stars with $M\le 2M_{\sun}$.  In particular, the recent
discovery of a compact star with a mass of 1.97~$M_{\sun}$ measured through the
Shapiro delay provides an observationally ``clean''
lower bound on the maximum mass of a compact
star~\citep{2010Natur.467.1081D}.

On the theoretical side it is now well-established that the emergence of
new degrees of freedom at high densities softens the equation of state
of matter. For example, allowing for the hyperons can reduce the
maximum mass of a sequence of compact stars below the canonical mass
of 1.4~$M_{\sun}$. A similar reduction may occur if a deconfinement to quark
matter takes place, although the softening of the equation of state in
this case is less dramatic. Thus, the observation of 2$M_{\sun}$ mass
neutron star is evidence that the ultra-dense matter in neutron
stars cannot be soft, \ie, agents that will substantially soften the
equation of state are potentially excluded.

We aim to study the equation of state of ultra-dense matter in the
light of this recent constraint~\citep{2010Natur.467.1081D}.  We
investigate to which extent one can reconcile the non-nucleonic
components such as hyperons and two- and three-flavor quark matter,
along with their color superconductivity, with the existence of
neutron stars with masses 2$M_{\sun}$.  The answer(s) to the question
above are of fundamental importance, because if hybrid stars featuring
quark cores surrounded by a (hyper)nuclear mantle exist in nature,
they could provide a unique window on the properties of quantum
chromodynamics (QCD) at high baryon densities under conditions not
attainable in laboratory experiments (the ultra-dense matter is in
equilibrium, is charge neutral and in $\beta$-equilibrium with respect
to weak interactions).

Although heavy baryons (mainly $\Sigma^{\pm}$ and $\Lambda$ hyperons)
were considered even before the discovery of pulsars and their
identification with the neutron stars~\citep{1960AZh....37..193A},
their emergence in the cores of neutron stars is still very
elusive. Treatments based on relativistic density functional
methods~\citep{1985ApJ...293..470G,1991PhRvL..67.2414G,weber_book}
predict masses that are not much larger than the canonical mass of a
neutron star which clearly contradicts  modern observations. Masses
on the order of $\le 1.8 M_{\sun}$ were obtained in non-relativistic
phenomenological
models~\citep{1999ApJS..121..515B,2010PhRvC..81c5803D},  while
microscopic models based on hyperon-nucleon potentials, which include 
the repulsive three-body forces, predict low maximal masses for
hypernuclear stars~\citep{PhysRev1998,2003astro.ph.12446B,2011EL.....9411002V}.

The treatment of deconfined matter at ultra-high densities is
model-dependent.  We used the Nambu--Jona-Lasinio (NJL) model to
describe the quark matter and its color superconductivity.  The model
is a non-perturbative low-energy approximation to QCD, which is
anchored in the low-energy phenomenology of the hadronic spectrum. The
dynamical symmetry breaking, by which quarks acquire mass, is
incorporated in this model, but it lacks confinement. Our ignorance of
the mechanism of confinement requires a free parameter in the theory,
which can be identified with the bag constant (the latter need not be
the same as in the MIT bag model). Furthermore, it can be eliminated
in favor of a more physical quantity - the transition density from
(hyper)nuclear to quark matter. In the low-density matter a candidate
superconducting phase is the two-superconducting-colors (2SC)
phase~\citep{1984PhR...107..325B}, which at sufficiently high
densities transforms to the three-flavor color-flavor-locked (CFL)
phase~\citep{1999NuPhB.537..443A}; our present study includes only
these two phases, but  we emphasize that the phase
structure of matter could be more
complicated~\citep{2005PhRvD..72c4004R}. The separate problem of
compact stars made of strange matter with equal number of flavors of
quarks~\citep{2011AIPC.1354...13W} will not be considered here.

Early studies of hadron-quark phase transition and quark
superconductivity within the NJL model suggested that no stable stars
(in the CFL phase) can be obtained within the standard
parameterization of this model (see, \eg,
~\citet{2003PhLB..562..153B}).  Recently, NJL-model based stable
sequences of hadron-quark stars were obtained for 2SC and
CFL matter~\citep{2006PhRvC..74c5802K,2007Natur.445E...7A} and a
three-flavor crystalline color superconducting
state~\citep{2008PhRvD..77b3004I}. The key requirement for the
existence (and stability) of these objects was the observation that the
nuclear equation of state must be stiff and the quark equation of
state must be supplemented by a shift (bag constant) to enable the
the quark and nuclear equations of
state to match~\citep{2008PhRvD..77b3004I}.  In the latter study stable
crystalline color superconducting stars with masse $M\sim 2M_{\sun}$
were obtained as ``twin'' configurations of purely nuclear
counterparts. Subsequently, sequences of stars containing homogeneous
quark matter in the 2SC and CFL phases (with maximum masses $\le 1.8
M_{\sun}$) were obtained in the NJL model supplemented by a repulsive
vector
interaction~\citep{2008PhRvD..77f3004P,2010PhRvD..81h5012L}. The
importance of the vector interactions in stiffening the quark equation
of state has been pointed out earlier by \citet{2006PhRvC..74c5802K}.
Furthermore, the emergence of twin configurations of
color-superconducting stars was confirmed in complementary studies
based on variations and extensions of the NLJ
model~\citep{2010JPhG...37i4063B,2010PThPS.186...81B,2010PhRvD..81b3009A}.

Massive hybrid configurations were also constructed within
phenomenological parameterizations of the quark matter equation of
state motivated by the MIT bag model in combination with nuclear
equations of state that are moderately soft; large masses require
strong quark-quark correlations in combination with color
superconductivity in the CFL
phase~\citep{2005ApJ...629..969A,2011arXiv1102.2869W}.

Below we shall adopt the point of view that the nuclear equation of
state needs to be stiff above the saturation density in order to
obtain massive hybrid stars~\citep{2008PhRvD..77b3004I}. With this
observation as a working hypothesis we will explore the range of
parameters allowing for compact stars featuring hyperonic and/or two-
and three-flavor quark matter.

\section{Models}

We start with the low-density part of the equation of state, where the
degrees of freedom are the nucleons and hyperons.  The nuclear
equation of state, as is well known, can be constructed starting from
a number of principles, see, \eg, \citet{weber_book} and
\citet{2007PrPNP..58..168S}. Below we will work with the relativistic
mean-field models, which are fitted to the bulk properties of nuclear
matter and hypernuclear data to describe the baryonic octet and its
interactions~\citep{1991PhRvL..67.2414G,1997PhRvC..55..540L} We adopt
the following Walecka Lagrangian, which includes self-interacting
$\sigma$-field
\begin{eqnarray}
\label{eq:L_RMF}
\mathcal{L}_{B}&=&\sum_B \bar{\psi}_B[\gamma^{\mu}(i\partial_{\mu}-g_{\omega B}\omega_{\mu
}-\frac{1}{2} g_{\rho B} {\bm \tau} \cdot {\bm \rho_{\mu}} )\nonumber\\
&-&(m_{B}-g_{\sigma B}\sigma)]\psi_B+\frac{1}{2}\partial^{\mu}\sigma\partial_{\mu
}\sigma-\frac{1}{2}m_{\sigma}^{2}\sigma^{2}
\nonumber\\
&+&\frac{1}{2}m_{\omega}^{2}\omega^{\mu}\omega_{\mu}-\frac{1}{4}{\bm
  \rho}^{\mu\nu} \cdot {\bm \rho}_{\mu\nu}+\frac{1}{2}m_{\rho}^{2}{\bm
  \rho}^{\mu}\cdot{\bm \rho}_{\mu}\nonumber\\
&-&\frac{1}{3}b m_{N} (g_{\sigma N} \sigma)^3
-\frac{1}{4}c (g_{\sigma N} \sigma)^4\nonumber\\
&+&\sum_{e^-,\mu^-}\bar{\psi}_{\lambda}(i
\gamma^{\mu} \partial_{\mu}-m_{\lambda})\psi_{\lambda}
-\frac{1}{4}F^{\mu\nu}F_{\mu\nu}
\text{,} 
\end{eqnarray}
where the $B$-sum is over the baryonic octet $B \equiv p, n, \Lambda,
\Sigma^{\pm,0}, \Xi^{-,0}$, $\psi_B$ are the corresponding Dirac
fields, whose interactions are mediated by the $\sigma$ scalar,
$\omega_{\mu}$ isoscalar-vector and $\rho_{\mu}$ isovector-vector
meson fields.  The meson and the baryon masses correspond to their
values in the vacuum, the values of nucleon-meson couplings are
$g_{\sigma N}/m_{\sigma}= 3.967 $ fm, $g_{\omega N}/m_{\omega}= 3.244
$ fm, $g_{\rho N}/m_{\rho}= 1.157 $ fm for nucleons; the hyperon-meson
couplings are obtained from these by multiplication by factors 0.6,
0.658, and 0.6, respectively. The couplings in the self-interaction
terms of the $\sigma$-field are given by $b= 0.002055 $ and $c =
-0.002651$. The next-to-last term in Eq.~(\ref{eq:L_RMF}) is the Dirac
Lagrangian of leptons, $F_{\mu\nu}$ is the energy and momentum tensor
of the electromagnetic field; we will not need the explicit from of
other tensors in the Lagrangian~(\ref{eq:L_RMF}). The parameters above
correspond to the NL3
parametrization~\citep{1997PhRvC..55..540L}. Computations were made
also with the GM3 parameterization~\citep{1991PhRvL..67.2414G}, and we
will comment on the differences below. The choice of this specific
parametrization was made because the nucleonic
matter has the stiffest equation of state compatible with the nuclear
phenomenology. The mean-field pressure of the (hyper)nuclear matter
can be obtained from Eq.~(\ref{eq:L_RMF}) in the standard
fashion~\citep{weber_book}.

The high-density quark matter is described by an  NJL Lagrangian, which
is extended to include the t' Hooft interaction term ($\propto K$) 
and the vector interactions ($\propto G_V$) 
\begin{eqnarray}
\label{eq:NJL_Lagrangian}
\mathcal{L}_{Q}&=&\bar\psi(i\gamma^{\mu}\partial_{\mu}-\hat m)\psi +G_V(\bar\psi i \gamma^{0}\psi)^2\nonumber\\
&+&G_S \sum_{a=0}^8 [(\bar\psi\lambda_a\psi)^2+(\bar\psi i\gamma_5 \lambda_a\psi)^2]\nonumber\\
&+& G_D \sum_{\gamma,c}[\bar\psi_{\alpha}^a i \gamma_5
\epsilon^{\alpha\beta\gamma}\epsilon_{abc}(\psi_C)^b_{\beta}][(\bar\psi_C)^r_{\rho} 
i \gamma_5\epsilon^{\rho\sigma\gamma}\epsilon_{rsc}\psi^8_{\sigma}]\nonumber\\
&-&K \left \{ {\rm det}_{f}[\bar\psi(1+\gamma_5)\psi]+{\rm det}_{f}[\bar\psi(1-\gamma_5)\psi]\right\},
\end{eqnarray}
where the quark spinor fields $\psi_{\alpha}^a$ carry color $a = r, g,
b$ and flavor ($\alpha= u, d, s$) indices, the matrix of quark current
masses is given by $\hat m= {\rm diag}_f(m_u, m_d, m_s)$, $\lambda_a$
with $ a = 1,..., 8$ are the Gell-Mann matrices in the color space,
and $\lambda_0=(2/3) {\bf 1_f}$.  The charge conjugate spinors are
defined as $\psi_C=C\bar\psi^T$ and $\bar\psi_C=\psi^T C$, where
$C=i\gamma^2\gamma^0$ is the charge conjugation matrix.  The partition
function of the system can be evaluated for the Lagrangian
(\ref{eq:NJL_Lagrangian}) neglecting the fluctuations beyond the
mean-field~\citep{2005PhRvD..72c4004R}.  To do so, one linearizes the
interaction term keeping the di-quark correlations $\Delta_c\propto
(\bar\psi_C)_{\alpha}^ai\gamma_5\epsilon^{\alpha\beta
  c}\epsilon_{abc}\psi_{\beta}^b$ and quark-anti-quark correlations
$\sigma_{\alpha}\propto\bar\psi_{\alpha}^a\psi_{\alpha}^a$. At zero
temperature the pressure is given by
\begin{eqnarray}
p&=&\frac{1}{2\pi^2}\sum_{i=1}^{18}\int_{0}^{\Lambda}dk k^2
\vert\epsilon_i\vert+4 K \sigma_u\sigma_d\sigma_s
\nonumber\\
&-&\frac{1}{4G_D}\sum_{c=1}^{3}\vert\Delta_c\vert^2
-2G_s\sum_{\alpha=1}^{3}\sigma_{\alpha}^2+\frac{1}{4
  G_V}(2\omega_0^2+\phi_0^2)\nonumber\\
&+&\sum_{l=e^-,\mu^-}p_l-p_0-B^*,
\end{eqnarray}
where $\epsilon_i$ are the quasiparticle spectra of quarks,
$\omega_0=G_V\langle QM \vert
\psi_u^{\dagger}\psi_u+\psi_d^{\dagger}\psi_d\vert QM\rangle$ and
$\phi_0=2 G_V\langle QM \vert \psi_s^{\dagger}\psi_s\vert QM\rangle$
are the mean field expectation values of the vector mesons $\omega$
and $\phi$ in quark matter, $p_l$ is lepton pressure, $p_0$ is the
vacuum pressure and $B^*$ is an effective bag constant.  The quark
chemical potentials are modified by the vector fields as follow
$\hat\mu^*={\rm diag}_f(\mu_u-\omega_0,\mu_d-\omega_0,\mu_s-\phi_0)$.
The numerical values of the parameters of the Lagrangian are $m_{u,d}
= 5.5$ MeV, $m_s = 140.7$ MeV, $\Lambda = 602.3$ MeV, $G_S\Lambda^2 =
1.835$, $K\Lambda^5 =12.36$, and $G_D/G_S = 1$.
\begin{figure*}[thb]
\centering
\resizebox{0.6\hsize}{!}{\includegraphics*{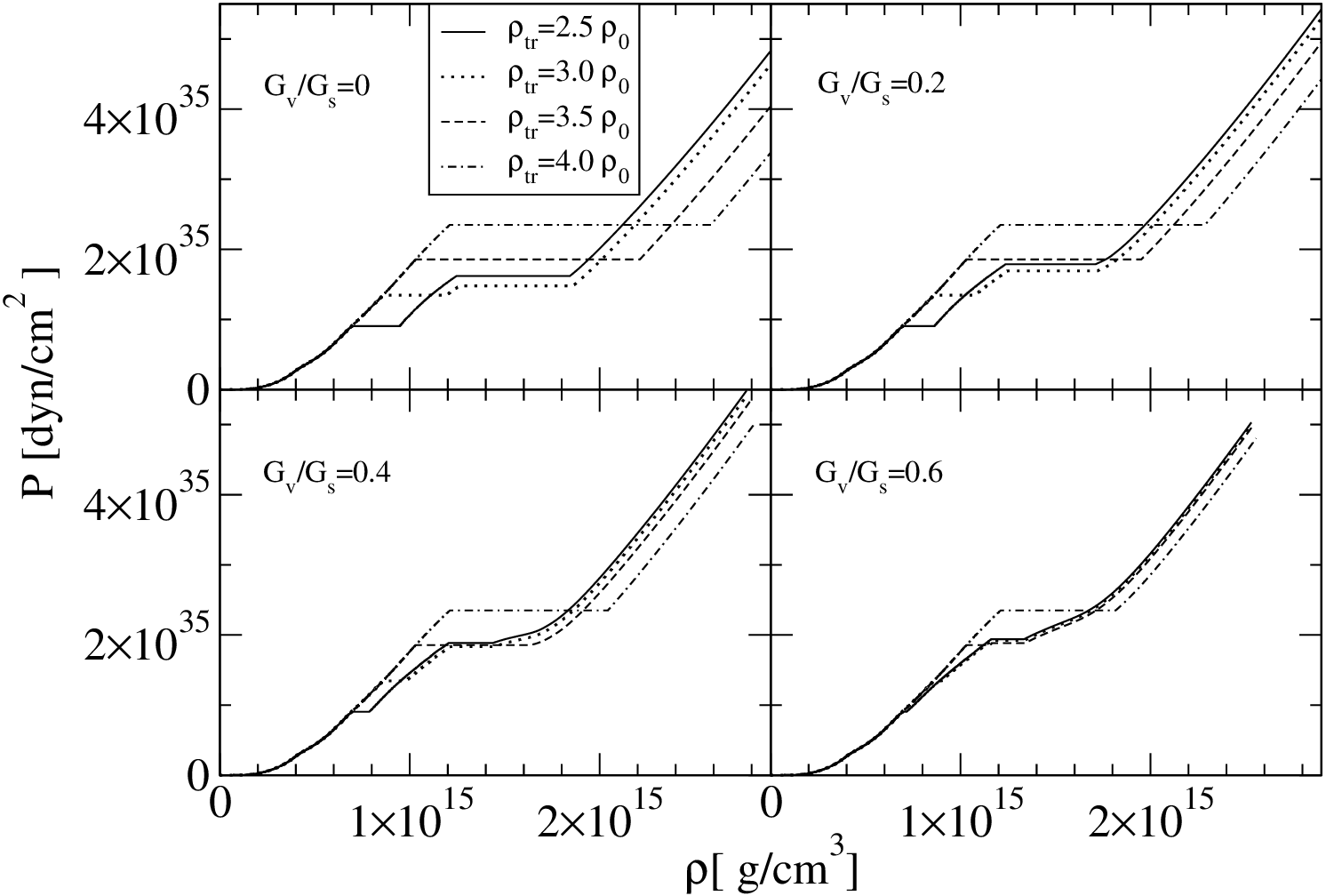}}\\
\caption{
Dependence of pressure on the density of matter in
$\beta$-equilibrium at $T=0$ for four values of transition density
from (hyper)nuclear matter to quark matter ($\rho_{\rm tr}/\rho_0=$ 2.5,
3, 3.5 and 4) .  The value of the vector coupling constant varies
from zero (upper left panel) to $0.6 G_S$ (lower right panel).
}
\label{eos}
\end{figure*}
The surface tension between the (hyper)nuclear and quark matter is not
well-known, therefore we shall adopt the working hypothesis that this
tension is high enough to prevent the formation of mixed phases. Then,
the transition from (hyper)nuclear matter to quark matter occurs at a
certain baryo-chemical potential at which the pressures of these
phases are equal. This is equivalent to the condition that pressure,
$P$, vs. chemical potential, $\mu$, curves for these phases cross
(Maxwell construction).   Thus, according to the Maxwell construction
of the deconfinement phase transition, there is a jump in the density
at constant pressure. However, the transition density itself cannot
be fixed, because the current NJL model does not allow us to fix the
low-density normalization of the pressure; (this is the consequence of
the fact that this class of models does not capture the confinement feature of
the QCD). For this reason we introduced above an additional ``bag''
parameter $B^*$, which allows us to vary the density at which the
quark phase sets-in, thus fixing the density of deconfinement
$\rho_{\rm tr}$.  The transition density increases with $B^*$ (as well as
with the vector coupling $G_V$). For example, varying $B^*$ in the
range -40 MeV fm$^{-3}$ to 50 MeV fm$^{-3}$ we find variations in the 
transition density in the range  $2.4\rho_0<\rho_{\rm tr}< 4 \rho_0$,
where  $\rho_0$ is the nuclear saturation density.

Before turning to the results of calculations, we briefly
summarize the physical input of the underlying models.  The masses and
coupling in the Lagrangian (\ref{eq:L_RMF}) are constrained by the
phenomenology of nuclear and hyper-nuclear matter; we will use the NL3
parameterization throughout.  The  parameters of the Lagrangian
(\ref{eq:NJL_Lagrangian}) are fixed to reproduce the 
observables of vacuum QCD: the pion mass $m_{\pi} = 135$ MeV, the kaon
mass $m_K = 497.7$ MeV, the eta-prime mass $m_{\eta'} = 957.8$ MeV and 
the pion-decay constant $f_{\pi} = 92.4$ MeV.  There remain two free
parameters $G_V$ and $B^*$. The first one can be used to regulate the
stiffness of the quark matter equation of state ($G_V=0$ gives the
softest equation of state), while $B^*$ can be varied to change the
density of the phase transition to the quark phase (deconfinement).
The contributions to the pressure from the condensates arises from two
phases: the two-flavor 2SC phase, where $\Delta_1 = \Delta_2 = 0$ and
$\Delta_3 \neq 0$ and the three-flavor CFL phase, where $\Delta_1 \neq
0$, $\Delta_2 \neq 0$, and $\Delta_3 \neq 0$. The strength of the pairing
field is fixed by setting $G_D = G_S$, which implies stronger pairing
interaction than the one deduced from the Fierz transformation 
$G_D = 0.75 G_S$.

\section{Results}

Figure~\ref{eos} shows the equations of state of (hyper)nuclear and
quark phases and their matching via Maxwell construction for several
values of the parameters $G_V$ and $\rho_{\rm tr}$. For low values
of the $G_V$ parameter there are two sequential transitions from
(hyper)nuclear matter to the 2SC phase and from 2SC phase to the CFL
phase.  As the onset density of the quark matter is shifted to higher
densities, the CFL phase is expelled from the density range relevant
for the phenomenology. For higher values of the $G_V$ parameter the
density jump(s) are smaller, therefore the CFL phase can nucleate at
lower densities.  This tendency eventually leads to a smooth
transition between the phases for $G_V = 0.6 G_S$, if the transition
density is not too high. The t' Hooft term
makes the transition from the 2SC to the CFL phase ``smoother''.

\begin{figure}[!]
\begin{center}
\includegraphics[width=8cm,height=7cm]{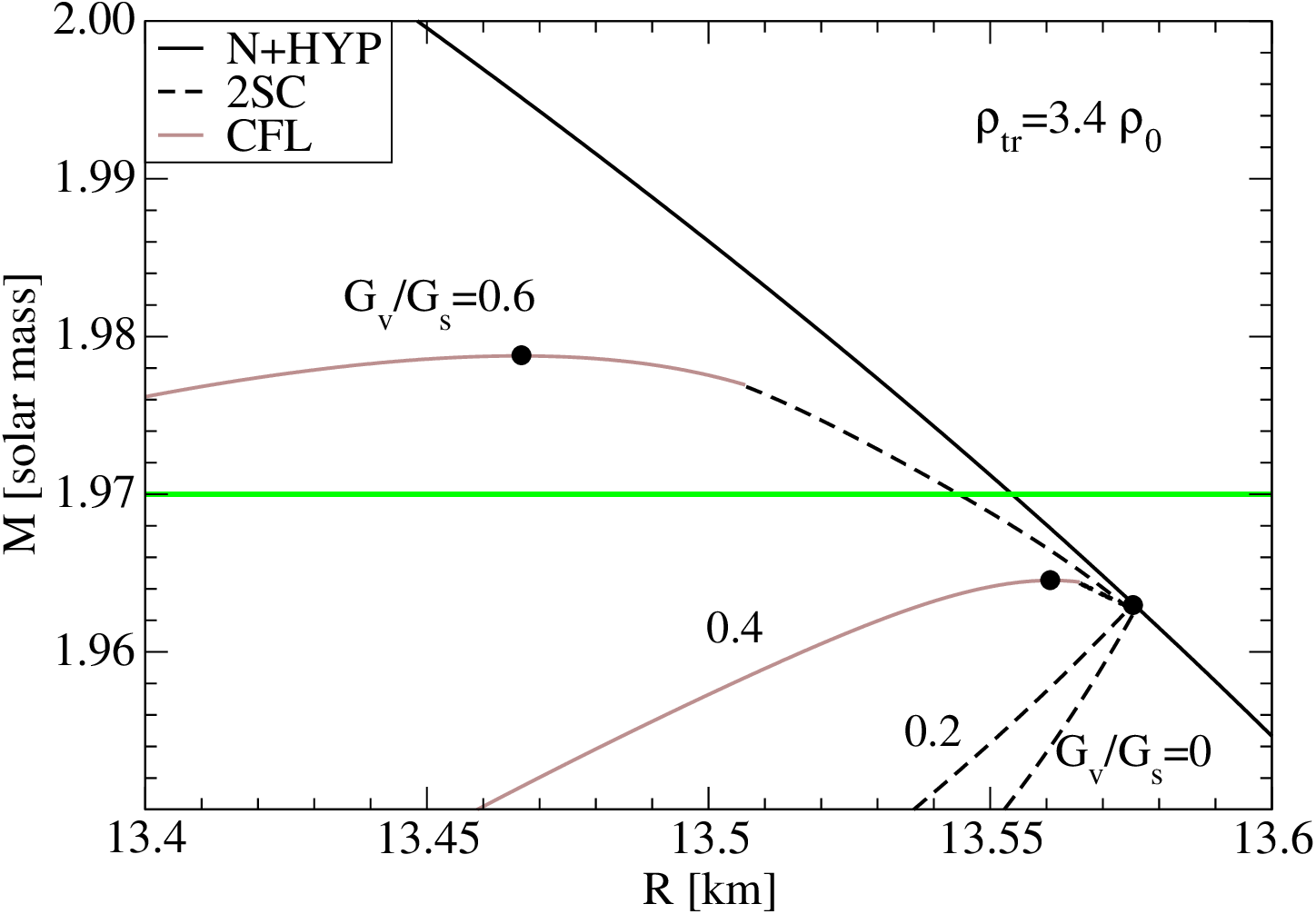}
\caption{Mass vs radius for configurations with quark-hadron
transition density $\rho_{\rm tr}=3.4\rho_0$ for four values of vector
coupling $G_V/G_S = 0, 0.2, 0.4, 0.6$. The purely hadronic sequence
(i.e. the sequence that includes nucleons and hyperons)
is shown by black solid line.  The dashed lines and the gray solid
lines show the branches where the 2SC and CFL quark phases are
present. The filled circles mark the maximum masses of the
sequences. The horizontal line shows the largest mass measurement to
date~\citep{2010Natur.467.1081D}.}
\label{fig:2}
\end{center}
\end{figure}

The equations of state described above were supplemented by an
equation of state of the low-density crust
\citep{1971ApJ...170..299B}.  These were used as an input into the
Oppenheimer-Volkoff equations to obtain sequences of stellar
configurations.  The resulting mass-radius relationship for massive
stars is shown in Fig.~\ref{fig:2} together with the largest mass
measurement to date $M= 1.97\pm 0.04 M_{\sun}$
\citep{2010Natur.467.1081D}.  Masses above the lower bound on the
maximum mass are obtained for purely hadronic stars; this feature is
prerequisite for finding similar stars with quark phases. 
Evidently only for high values of vector coupling $G_V$ one finds stable
stars that contain (at the bifurcation from the hadronic sequence)
the 2SC phase, which are followed by stars that additionally contain
the CFL phase (for higher central densities).  Thus we find that the
stable branch of the sequence contains stars with quark matter in the
2SC and CFL phases. Similar results were obtained from the GM3
parameterization of nuclear matter if, however, the hyperons were
excluded from the consideration. Therefore, a less stiff GM3 parameterization
is qualitatively equivalent to the NL3 parametrization, which features
the softening of the equation of state due to the hyperons.

\begin{figure}[htb!]
\begin{center}
\includegraphics[width=8cm,height=7cm]{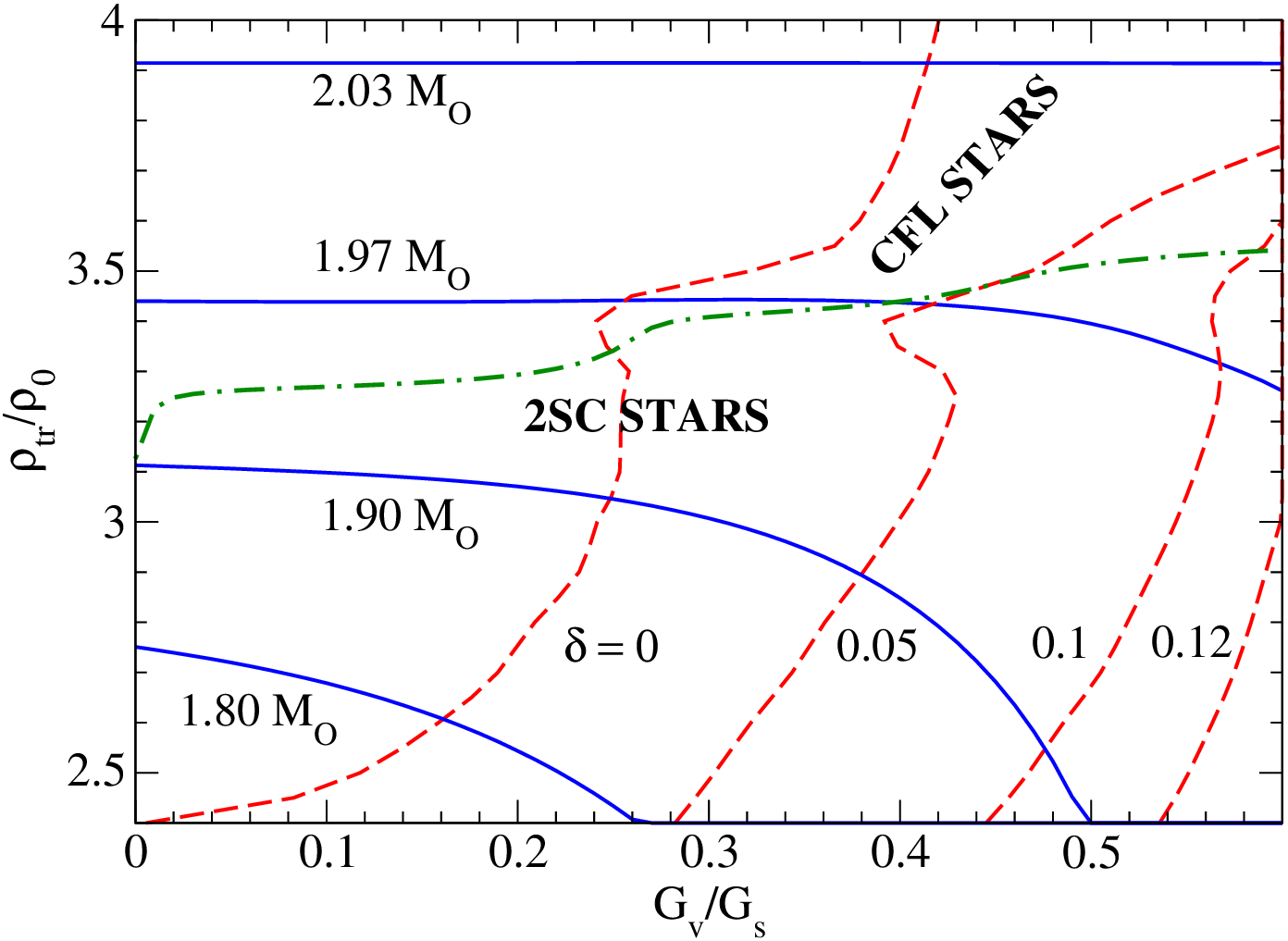}
\caption{Properties of the stars as a function the free parameters
$G_V$ and $\rho_{\rm tr}$. The solid lines (blue online) show the
maximum mass configurations realized for the pair of parameters
$G_V$ and $\rho_{\rm tr}$. The dashed (red online) curves show the
amount of CFL matter in the configurations via the ratio $\delta =
R_{CFL}/R$, where $R_{CFL}$ is the radius of the CFL core, $R$ is
the star radius. The parameter space to the right from $\delta = 0$
line produces CFL stars.  The parameter space below the
dashed-dotted (green online) curve corresponds to stars containing 
2SC matter.
}
\label{fig:3}
\end{center}
\end{figure}

Figure \ref{fig:3} summarizes several results for the masses and
composition of compact stars as a function of the parameters of the
model $G_V$ and $\rho_{\rm tr}$. First, it shows the tracks of
constant maximum mass compact stars within the parameter space. The
decrease of these lines with increasing vector coupling reflects the
fact that non-zero vector coupling stiffens the equation state. In
other words, to obtain a given maximum mass one can admit a small
amount of soft quark matter with vanishing vector coupling by choosing
a high transition density; the same result is obtained with a low
transition density, but strong vector coupling, \ie, a stiffer quark
equation of state. For low transition densities one finds 2SC matter
in stars, which means that weaker vector couplings slightly disfavor 2SC
matter. Substantial CFL cores appear in configurations for strong
vector coupling and almost independent of the transition density
(nearly vertical dashed lines with $\delta\sim 0.1$ in
Fig. \ref{fig:3}).  Note that for a high transition density there is a
direct transition from hyper-nuclear to the CFL phase. For transition
densities blow $3.5\rho_0$ a 2SC layer emerges that separates these
phases. On the other hand, weak vector couplings and low transition
densities produce stars with a 2SC phase only.
\begin{figure}[htb]
\begin{center}
\includegraphics[width=8cm,height=7cm]{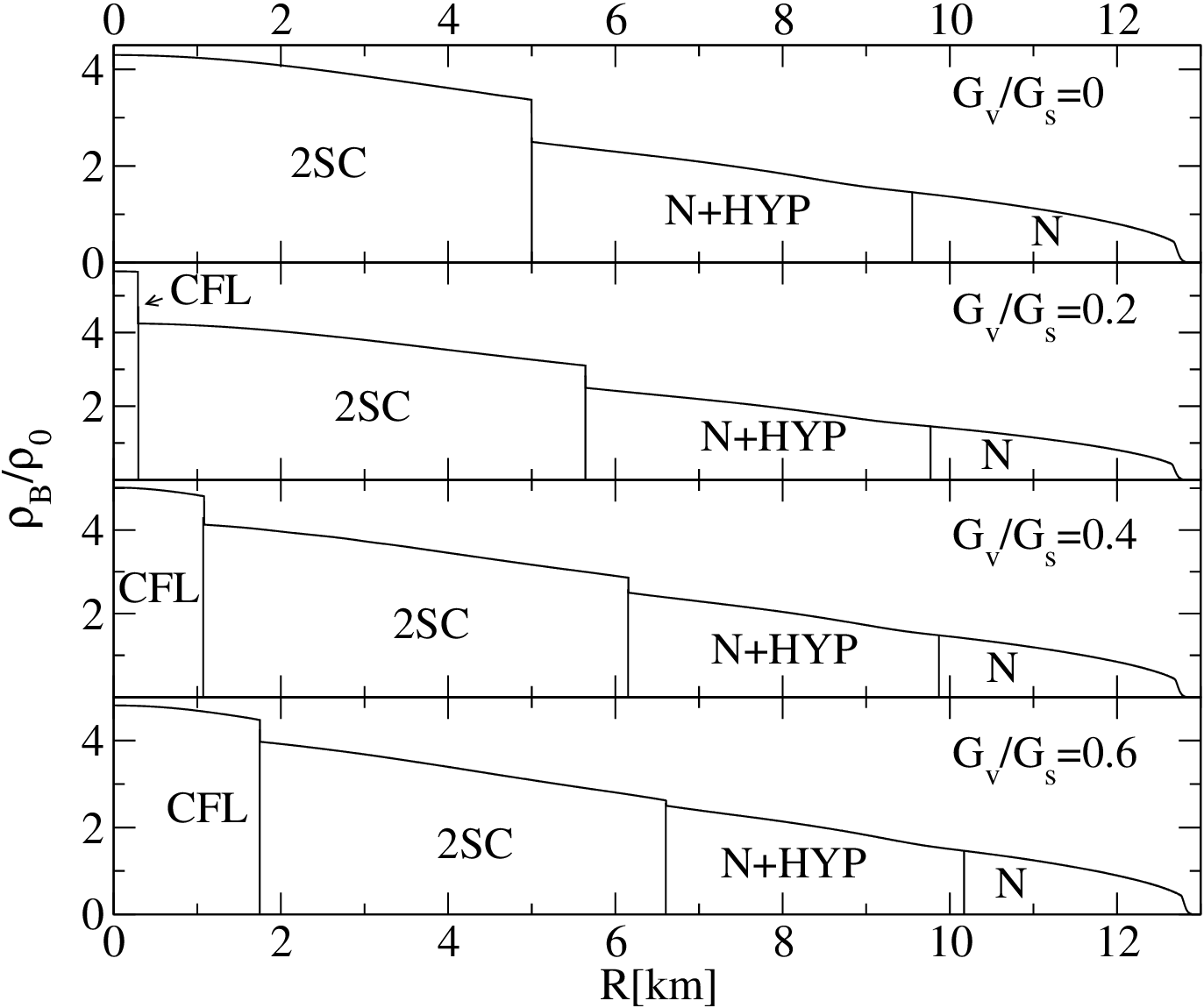}
\caption{Density profiles of the maximum mass configurations with the
quark-hadron transition density $\rho_{\rm tr}=2.5 \rho_0$. The
vector coupling constant varies from zero (upper panel) to $0.6 G_S$
(lower panel).  The hyper-nuclear and nuclear phases are marked by
N+HYP and N respectively.  }
\label{fig:4}
\end{center}
\end{figure}

In Fig. \ref{fig:4} we show the baryon density profiles of the
maximum-mass stars with a transition density of $\rho_{\rm tr}=2.5
\rho_0$. For soft quark matter $G_V=0$ we obtain quark matter only in
the 2SC phase, which extends up to about 1/3 of the star
radius. Increasing $G_V$ has the effect of shifting the 2SC phase to
larger radii, whereas the CFL phase develops at the center. For the
strongest coupling studied, all phases are present in the maximum
mass star, with the combined 2SC and CFL paired quark matter confined within a radius
that is about the half of the star. Note that a similar internal structure is
obtained for crystalline color-superconducting stars, where quark
matter is confined within a radius of $\sim 7$~km for maximum mass
stars with a radius $\sim 12$~km~\citep{2009PhRvD..79h3007K}.

\section{Conclusion}

Massive neutron stars are likely to develop cores composed of
deconfined quark matter, which should be in one of the color
superconducting phases. We have constructed an equation of state of
such matter on the basis of relativistic mean-field nuclear functional
at low densities and effective NJL model of quark matter supplemented
by the t' Hooft and vector interactions. Non-rotating spherically
symmetric configuration with maximum masses $\sim 2 M_{\sun}$ were
studied by solving the Oppenheimer-Volkoff equations and were compared
to the recent lower bound on the maximum mass of a neutron star. Our
equation of state contains two free parameters: the transition density
from (hyper)nuclear matter to quark matter and the vector coupling of
quarks.  We have demonstrated that stable configurations featuring
color superconducting matter can be constructed if vector interactions
are included in the quark equation of state. A prerequisite for this
is a stiff nuclear equation of state. We find that the NL3
parameterization with hyperons or GM3 parameterization with nucleons
only are sufficiently stiff to produce maximum masses above the lower
bound. Inclusion of quark degrees of freedom softens these equations
of state; the required lower bound can still be achieved because the
vector interactions in the quark matter can stiffen the quark equation
of state. 

To conclude, relativistic hypernuclear Lagrangians, which predict
stiff hypernuclear equations of state above saturation density, enable
us to construct stable configurations with masses equal and above the
measured 1.97 solar-mass star. Our configuration have ``exotic''
matter in their interiors in the form of hyperons and quark matter,
of which the quark matter is color-superconducting in the two-flavor
2SC and/or three-flavor CFL phases. The choice of parameters in our
models geared toward producing stiff equations of states above
the saturation density. Observations of pulsars with larger masses
(above $2M_{\sun}$) will seriously challenge our current understanding
of dense matter physics above nuclear saturation.

\section*{Acknowledgements}
This work was supported in part by the Helmholtz International Center
for FAIR within the framework of the LOEWE (Landesoffensive zur
Entwicklung Wissenschaftlich-\"Okonomischer Exzellenz) program
launched by the State of Hesse and by the Research Networking
Programme of the European Science Foundation {\it ``Compstar''}.

\end{document}